\documentclass[aps,twocolumn,preprintnumbers, amsmath, amssymb, floatfix]{revtex4}
\usepackage{dcolumn}
\usepackage{bm}

\def\be{\begin{equation}}
\def\ee{\end{equation}}
\def\beq{\begin{eqnarray}}
\def\eeq{\end{eqnarray}}
\def\n{\nonumber}

\begin{document}
\preprint{smw-ahw-06-L01}
\title{The Gravity and The Quantum: A Bohr-inspired Synthesis}

\author{Sanjay M. Wagh}
\affiliation{Central India Research Institute, Post Box 606,
Laxminagar, Nagpur 440 022, India}
\email{cirinag_ngp@sancharnet.in}

\author{Abhijit H Wagh}
\affiliation{B.E. II-nd Year Student, Birla Institute of
Technology \& Science, Pilani 333 031, India}
\email{jetwagh@yahoo.co.in}

\date{August 2, 2006}
\begin{abstract}
An effective angular momentum quantization condition of the form
$mvr=n\hbar(m/m_F)$ is used to obtain a Bohr-like model of
Hydrogen-type atoms and a modified Schr\"{o}dinger equation.
Newton's constant, $G$, of Gravitation gets explicitly involved
through the fundamental mass $m_F$ as defined in the sequel. This
non-relativistic formalism may be looked upon as a ``testing
ground'' for the more general synthesis of the gravity and the
quantum.
\end{abstract}
\maketitle

One of the pivotal principles of physical nature underlying
mathematical methods of the quantum theory can be said to be the
following.

Quantization of a physical quantity is, at a very basic level,
expressing that physical quantity as an integral multiple of
``some unit'' of that quantity, that unit being obtainable from
the fundamental constants of Nature alone. This principle based on
the dimensional analysis has a crucial role in not only the
orthodox quantum theory but also in subsequent developments in
Physics.

As is well known, Niels Bohr, in a masterly use of this principle,
had postulated the quantization rule for the angular momentum of
an electron (of mass $m_e$ and charge $e$) orbiting an atomic
nucleus (of atomic number $Z$) in his famous model of the hydrogen
atom as \cite{qmech}:
\[ m_ev_nr_n = n\hbar \] to express the angular
momentum as an integer, $n$, multiple of Planck's constant,
$\hbar$, a fundamental constant. As is also well known, having
imposed this condition, the wave number of a photon that is
emitted by the atom transiting from a state $n_2$ to a state $n_1$
($n_2 > n_1$) is given by
\[ \frac{1}{\lambda} = Z^2R \left[
\frac{1}{n_1^2} - \frac{1}{n_2^2} \right] \] where $R =e^4m_e/
2\hbar^2c$ ($c$ being the speed of light in vacuum) is the Rydberg
constant, which now gets determined completely in terms of the
fundamental constants of Nature.

The force of gravity between an atomic nucleus and an electron is
about $10^{-42}$ times weaker than their electrostatic force of
attraction. Still, it does seem somewhat unsatisfactory that
Bohr's model does not incorporate basic constants other than
Planck's constant, like Newton's constant $G$ of gravitation, a
fundamental constant.

If gravity and the quantum nature of the physical phenomena are to
result from only a single theory, then it seems plausible that we
will have some limited formulation of the quantum theory and,
hence, of Bohr-like model of the atom in which $G$ would also
appear along with $\hbar$ and other fundamental constants of
Nature. However, no such formulation is currently available.

Perhaps, this lacuna can be remedied by using a ``unit'' for the
angular momentum different than $\hbar$ that was used by Bohr.
Needless to say, but we still wish to emphasize at this place
that, any new unit of the angular momentum must be constructible
out of the basic constants.

In this paper, we show that this can indeed be done. We also
provide a Schr\"{o}dinger-like equation for the formulation of
corresponding ideas. Just as Bohr's model provided the testing
ground for developments in the orthodox quantum theory, the
proposed Bohr-like model may be a testing ground for the general
synthesis ({\em eg}, \cite{smw-utr, smw-let-0601,
smw-sars,smw-ppt, lir-gr}) of the gravity and the quantum.

To begin with, we note that using $c$, $h$, $G$, $m_e$ and $e$, we
can form a quantity that has the dimensions of the angular
momentum as: \be A_P=e^2\,m_e\,\sqrt{\frac{G}{c^3h}} \approx 4.05
\times 10^{-53}\; \mathrm{cgs\;units}. \label{angunit} \ee It is
however very tiny. Notably, Planck's constant $\hbar$ is $2.61
\times 10^{25}$ times larger than it.

Therefore, to be able to express the ``usual'' angular momentum of
an atomic electron in terms of $A_P$, we ``scale'' this tiny unit
by a factor $\Upsilon$ and express Bohr's condition of the
quantization of the angular momentum as

\be m_ev_nr_n = n\Upsilon A_P = n\;\Upsilon
e^2\,m_e\,\sqrt{\frac{G}{c^3h}} \label{angcon} \ee

We could strongly object to the occurrences of $e$ and $m_e$ in
(\ref{angunit}) since these are not fundamental constants of
Nature as are $G$, $c$ and $h$, but are particle-specific
quantities. However, we also note that (\ref{angcon}) is
equivalent to expressing the linear momentum of an atomic electron
as: \be m_ev_n = \frac{Z}{n\Upsilon}\;\sqrt{\frac{c^3h}{G}}
\label{momcon} \ee where the square-root is related to Planck's
length $L_p=\sqrt{hG/c^3}$ by $h/L_p$.

Evidently, $e$ and $m_e$ do not at all play any role in condition
(\ref{momcon}). Essentially, the quantization condition for one
physical quantity implies the quantization of another related
physical quantity with the proportionality factors of one
affecting those of the other \footnote{For Bohr's model using
Planck's constant as a unit of the angular momentum, the linear
momentum is given as $m_ev_n = Ze^2m_e/n\hbar=(Z/n)\alpha m_e c$.
Proportionality factors have been adjusted to get $Z/n$ for the
momentum in (\ref{momcon}). It is then that the formula for the
wave number of emitted radiation has the form of
(\ref{wnumber}).}.

In (\ref{momcon}), we have chosen proportionality constants to
``match'' the formulas of the present study with those of the
``usual'' atomic physics. Thus, the factor of $Z/n\Upsilon$ is
``adjustable'' in the condition (\ref{momcon}), we may then note
here.

It then follows that the radius of the electronic orbit of the
atom is given by \be r_n = \frac{n^2\Upsilon^2}{Z}
\frac{e^2m_eG}{c^3h} \label{atomrad} \ee and that the wave number
of a photon, which is emitted by the atom transiting from a state
$n_2$ to a state $n_1$ ($n_2> n_1$), is given by: \be
\frac{1}{\lambda} = \frac{Z^2c^2}{2\Upsilon^2Gm_e} \left[
\frac{1}{n_1^2} - \frac{1}{n_2^2} \right] \label{wnumber} \ee

It now immediately follows that, for an appropriate value of
$\Upsilon$, the same atomic physics would result even with the
newly chosen unit of the angular momentum.

On its face value, the above exercise appears to be only of pure
academic interest and, hence, not useful for further progress, if
the same atomic physics as that of Bohr's model is to result from
it. But, we would like to note that the validity of the proposed
model can be beyond that of Bohr's model using Planck's constant
simply because $\Upsilon$ is a free parameter.

We also note that the validity condition for the non-relativistic
treatment reads:
\[ v_n/c \ll 1 \Rightarrow \;\Upsilon^{-1} \ll \frac{nm}{Z}
\sqrt{\frac{G}{ch}} \approx \frac{nm}{Z} \times 5.6\times 10^3
\] which, for an electron revolving around the usual atomic
nucleus, is
\[ v_n/c \ll 1 \Rightarrow \;\Upsilon^{-1} \ll \frac{n}{Z}\times 5.1\times
10^{-23} \]

For comparison with Bohr's model, we note that the condition for
non-relativistic treatment reads $(Z/n)\,\alpha << 1$ and that
this is independent of the mass of the particle under
consideration \footnote{This is what leads us to doubt the
validity of the non-relativistic treatment for ``strange atoms''
like $\mu$-meson atoms , as have been considered in many
textbooks. Our work here implies however that the non-relativistic
treatment can be applicable in certain of such cases.}. The first
of our inequalities, with mass, can be satisfied for suitable
values of $m$, $e$ and $\Upsilon$.

After having demonstrated the plausibility of the Bohr-like model
\footnote{AHW worked out this Bohr-like model following the
suggestion of SMW of exploring the changes to Bohr's model with
the condition (\ref{momcon}) in the form $m_ev_n = \beta
\sqrt{c^3h/G}$ with $\beta$ as a dimensionless proportionality
constant.} of the atom involving the new unit of angular momentum
or, equivalently, the new unit of momentum, we turn to aspects
related to ideas of Schr\"{o}dinger.

Now, as is well known \cite{qmech}, Schr\"{o}dinger's equation is
obtained from $E=p^2/2m + V$, where $p$ is the momentum and $V$ is
the potential, by substituting $p\to -\imath\hbar\nabla$ and $E\to
\imath\hbar \partial/\partial t$ wherein Planck's constant $\hbar$
serves to keep the dimensions of the terms correct. (All the
``undeclared'' symbols will have their usual meaning.)

In the above spirit, we now use the ``new unit'' (that has the
dimensions of Planck's constant) and make the obvious
substitutions $p\to -\imath A_P \nabla$ and $E\to \imath A_P
\partial/\partial t$ in $E=p^2/2m_e + V$. This leads us to the
following Schr\"{o}dinger-like equation: \beq -\;\frac{\Upsilon
e^2m_e}{\imath}\,\sqrt{\frac{G}{c^3h}}
\frac{\partial\Psi}{\partial t} = &-&\frac{e^4m_e\Upsilon^2}{2}\,
\frac{G}{c^3h}\;\nabla^2\Psi \n \\ &+& V\Psi \n \eeq

For a free particle, this equation reduces to
\[ \imath \frac{\partial\Psi}{\partial t} = - \;
\frac{e^2\Upsilon}{2}\,\sqrt{\frac{G}{c^3h}}\; \nabla^2\Psi \]
Notice that $m_e$ drops out of the equation. The factor $\Upsilon$
then essentially determines the ``fundamental'' mass, $m_F$, as
\beq m_F = \frac{1}{2\pi \Upsilon e^2}\,\sqrt{\frac{c^3h^3}{G}}
&=& \frac{1}{\alpha \Upsilon}\,\sqrt{\frac{ch}{G}}\n \\ &=&
\frac{0.02374}{\Upsilon} \;\mathrm{gm} \label{fmass} \eeq  where
$\alpha=e^2/c\hbar\approx 1/137$ is the well known fine structure
constant.

With this definition, the Schr\"{o}dinger equation given above can
be written as: \be -\imath \hbar \left( \frac{m}{m_F}\right)
\frac{\partial\Psi}{\partial t} = -\frac{\hbar^2}{2m} \left(
\frac{m}{m_F}\right)^2 \nabla^2\Psi + V\Psi \label{scheq} \ee
wherein we dropped the subscript $e$ on the mass $m$, as this
equation holds generally.

With the definition (\ref{fmass}), the Bohr quantization condition
(\ref{angcon}) then reads $m_ev_nr_n = n\hbar (m_e/m_F)$, while
the quantization condition (\ref{momcon}) reads $m_ev_n =
(Z/n)\alpha m_F c=(Z/n)(e^2m_F/\hbar)$. It should then be noticed
that we only have ``scaled'', by $m/m_F$, Planck's constant of the
usual Schr\"{o}dinger's equation, and also that no new solutions
are implied here. Clearly, standard solutions of Schr\"{o}dinger's
equation can be used to explore the consequences of the existence
of $m_F$ in (\ref{scheq}).

We have then obtained here, only a preliminary, Bohr-inspired,
synthesis of the gravity and the quantum by way of the fundamental
mass $m_F$ and the Schr\"{o}dinger equation (\ref{scheq}) at the
non-relativistic level of its applications.

As closing remarks, one of us (SMW) would like to comment as
follows.

Evidently, we will recover,  as was stated before, the ``usual''
atomic physics when $m=m_F=m_e$. (In this case, $\Upsilon=(\alpha
m_e)^{-1}\,\sqrt{ch/G}$.) The proposed preliminary synthesis of
gravity with the quantum then incorporates the usual explanations
of quantum phenomena.

An apparent limitation of (\ref{fmass}) and, hence, of
(\ref{scheq}) is that these involve charge, {\em ie}, in
(\ref{fmass}), $e\neq 0$. But, this is due to the following
situation.

Notice here that $c^3/GX^2$ has the dimensions of Planck's
constant if $X$ has the dimensions of $1/L$, and $1/L$ are the
dimensions of Rydberg's constant. Thus, no quantity of the
dimensions of angular momentum, other than $\hbar$, can be
constructed from $G$ and $c$ without involving $\hbar$ or any
particle-specific quantities like charge and mass.

Now, momentum has the dimensions of $[\hbar]/L$. Hence, similar to
the above case with the angular momentum,  no quantity having
dimensions of momentum can be constructed out of $G$ and $c$
without explicitly involving $\hbar$ or any particle-specific
quantities like charge and mass.

But, $e^2m_F$ is completely determined by fundamental constants
$G$, $c$, $\hbar$, and the electronic charge $e$ appears in the
form of the fine structure constant only. Thus, the appearance of
$e$ in (\ref{fmass}) is not any serious limitation.

Then, the appearance of $e$ in (\ref{fmass}) rather points at a
way in which equations of a more general formulation could be
expected to reduce to their non-relativistic forms. Of course,
unless a satisfactory mathematical framework is at hand for the
general synthesis of gravity and the quantum, it is difficult to
elaborate on as to how this reduction of general equations to
(\ref{scheq}) obtains.

It is however clear that we have only two possibilities in the
non-relativistic domain: either the usual Schr\"{o}dinger equation
or the equation (\ref{scheq}). As stated before, equation
(\ref{scheq}) is ``equivalent'' to the usual Schr\"{o}dinger's
equation, albeit with ``modified'' Planck's constant. The modified
Planck's constant will arise unless of course $m=m_F$, {\em ie},
if $\Upsilon =(\alpha m)^{-1}\,\sqrt{ch/G}$ always. In this last
case, the onus then falls on the more general formulation of the
synthesis of the gravity with the quantum to explain as to why an
entirely arbitrary constant $\Upsilon$ always has this form!

As this last possibility appears preposterous, we look upon an
equation of the form (\ref{scheq}) as the ``proper'' quantum
equation of the non-relativistic domain in that equations of a
more general formulation should reduce to (\ref{scheq}). The
following is supportive of this view.

We then notice that if the fundamental mass $m_F$ were taken to be
dependent on different powers of the fine structure constant
$\alpha$ by demanding that $\Upsilon \propto 1/\alpha^N$ ($N$ - an
integer), {\em ie}, if $m_F = \alpha^{N-1} \sqrt{ch/G}$, then we
obtain a spectrum of fundamental mass values for the chosen value
of charge $e$. (Different such powers of $\alpha$ may be expected
to arise in a synthesis of gravity with the quantum more general
than the present one.)

Another possibility is of an integral multiple of the fundamental
mass $m_F$ as a possible origin of the mass spectrum. In this
case, $\Upsilon$ could be chosen to yield the right mass for one
particle, say, $\pi^o$-meson. Then, the observed \cite{epmass}
mass spectrum of some of the elementary particles is
``explainable'' within the present formalism.

The so-obtained mass spectra should then turn out be relevant for
the mass spectrum of elementary particles \cite{epmass}. We would
like to note however that the equation (\ref{scheq}) does not
provide any ``explanation'' for the masses of elementary particles
or even the mass spectrum.

In this connection, we would like to also note that important
effects of spin and other characteristics of elementary particles
are not included in the present simple analysis. The
aforementioned mass spectrum should therefore be looked upon only
as a feature of the proposed synthesis. As compared to the usual
quantum formulation, the equation (\ref{scheq}) then has quite a
``direct'' relevance to the non-relativistic behavior of each
member of the aforementioned mass spectrum.

Now, this synthesis of the gravity and the quantum can be expected
to have observable consequences. In particular, interesting
situations \footnote{We propose to explore different such
observable consequences of $m_F$ in a later work.} could be
expected to arise when $m\neq m_F$ in (\ref{scheq}). We then look
for ``quantum'' situations, {\em eg}, those considered in
\cite{chiao}, wherein the ``effective'' value of Planck's constant
could be ``different'' than that in, {\em eg}, Planck's radiation
formula.

Of some particular interest in this connection is the phenomenon
of Bloch oscillations \cite{harrison, kittle} of electrons in a
crystal. Although it has been difficult to observe Bloch
oscillations in natural crystals, recent advances in laser
spectroscopy have made it possible to observe  \cite{ferrari} the
Bloch oscillations of ultra-cold atoms in optical lattices
generated by interfering laser beams trapping atoms.

For laser generated optical lattices, the Bloch frequency is given
by $\nu_B=mg\lambda/2h$ where $\lambda$ is the frequency of laser
light producing the interference that traps atoms of mass $m$ in a
uniform gravitational field of acceleration $g$. (This is based on
standard Schr\"{o}dinger's equation.) It should then be possible
to ``create'' situations wherein the ``effective'' Planck's
constant should be experimentally observable confirming or
refuting thereby the proposed synthesis of gravity and the quantum
at the non-relativistic level.

We would like to reemphasize that any non-observation of the
effective Planck's constant, as implied by the present study, will
be a certain theoretical surprise: we will have to understand then
as to why Nature has selected an arbitrary constant $\Upsilon$
only in a particular way.

Lastly, we would like to note the following. Any conceptual and
mathematical framework of the unification of all the fundamental
interactions will have to necessarily incorporate all the
fundamental physical constants along with particle-specific
physical variables such as mass and charge of the particle. In
suitable approximation, whose manner is unclear at the present, we
will have to recover the equation (\ref{scheq}) from the
mathematical formalism of the theory of unification. This is of
course subject to the aforementioned caveat that we infer the
existence of the effective Planck's constant in some experimental
situation.

Clearly, this preliminary synthesis of the gravity with the
quantum already has consequences for general physical situations
and for a fundamental theory of the unification of all the
physical interactions, both. Although it is quite an elementary
and, hence, limited formalism, its potential implications appear
to be fundamental to further progress. It would therefore be
important to experimentally establish the existence of an
effective Planck's constant as in (\ref{scheq}).

\end{document}